\documentclass[preprint,12pt]{elsarticle}

\usepackage{graphicx}
\usepackage{epsfig}

\usepackage{amssymb}

\journal{Physics Letters A}

\begin{document}
\begin{frontmatter}
\title{Nonequilibrium thermal entanglement for three-spin chain}
\author{N. Pumulo}
\author{I. Sinayskiy \corref{cor1}}
\cortext[cor1]{Corresponding author. Tel./Fax: +27-(0)31-260-8133/8090\\ e-mail address: sinayskiy@ukzn.ac.za}
\author{F. Petruccione}
\address{Quantum Research Group, School of Physics and National
Institute for Theoretical Physics, University of KwaZulu-Natal,
Durban, 4001, South Africa}

\begin{abstract}
The dynamics of a chain of three spins coupled at both ends to
separate bosonic baths at different temperatures is studied.  An
exact analytical solution of the master equation in the
Born-Markov approximation for the reduced density matrix of the
chain is constructed. It is shown that for long times the reduced
density matrix converges to the non-equilibrium steady-state.
Dynamical and steady state properties of the concurrence between
the first and the last spin are studied. 
\end{abstract}


\end{frontmatter}

\section{Introduction}
In the description of a real physical system effects of the
environment play an important role \cite{TOQS}. Typically, the
interaction with the surroundings destroys quantum correlations in
the system. However, in some situations, interaction with the
environment can create extra quantum correlations in the system
\cite{braun}. Over the past few years the phenomenon of thermal
entanglement has been extensively studied
\cite{TE,SinBurPet,Huang,Qui}. When the quantum system is in
contact with a thermal reservoir, after a certain time the system
relaxes into its thermal state in a canonical distribution form
and thermal entanglement for such a scenario has been also widely
studied in the past few years \cite{TE}. A more general situation arises if the quantum system is coupled to two thermal reservoirs at different temperatures. To model the quantum analog of the Fourier law one natural choice is to consider a chain of spins connected at the ends to thermal reservoirs at different temperatures \cite{FL}.

In a previous paper
\cite{SinBurPet} thermal entanglement which is created in a two
spin system in contact with two heat reservoirs at different
temperatures was studied. Huang et al. \cite{Huang} studied
numerically a non-equilibrium three spin system. The model investigated in this paper is slightly simpler than the
one studied by Huang et al. numerically \cite{Huang}. We will
assume, as Huang et al. did, a chain of spins coupled by XX
interaction. Huang et al. consider the possibility of different
strength of the spin-spin interaction between the first and the
second spin and the second and the third one. Here, we assume
that the strength of all spin-spin interactions is the same. This
assumption will allow us to construct an exact analytical
expression for the density matrix of the three spins system. We
will show that the reduced density matrix of the three spin chain
is converging with time to the steady state.

The analytical solution for the three spin system is compared to
the corresponding solution of a two spin model, that was studied
previously \cite{SinBurPet}. Calculation of the concurrence
between extreme spins in the three spin chain and concurrence in
the two spin chain, reveals that adding and intermediate spin into
the chain increases steady state concurrence for a certain range
of temperatures of the baths.

The paper is organized as follows. In Section 2 we introduce the
model of a chain of three spins coupled at both ends to bosonic
baths at different temperatures. The model is a natural
generalization of the one studied in \cite{SinBurPet}. For
completeness we follow \cite{SinBurPet} in deriving the master
equation for the reduced density matrix of the spin system in the
Born-Markov approximation. Thus, the master equation has a
structure similar to the one in \cite{SinBurPet}. In Section 3 we
present the analytical solution and show the convergence of the
obtained solution to the non-equilibrium steady-state. Unlike in
the two spin chain case \cite{SinBurPet}, in the three spin chain
case the time dependence of the non-diagonal elements has a more
complicated form and we show analytically that all non-diagonal
elements vanish with time. Finally, in Section 4 we present
results and conclude.

\section{The model}
We consider a spin chain consisting of three spins. The first and
the last spin are coupled to separate bosonic baths at different
temperatures. The total Hamiltonian of the system is given by

\begin{equation}
\hat{H}=\hat{H}_S+\hat{H}_{B1}+\hat{H}_{B3}+\hat{H}_{SB1}+\hat{H}_{SB3},
\end{equation}
where $\hat{H}_S$ is the Hamiltonian of the spin subsystem,
\begin{equation}
\hat{H}_S=\frac{\epsilon}{2}\sum_{i=1}^3\hat{\sigma}^z_i+\kappa\sum_{i=1}^{2}\left(\hat{\sigma}^+_i\hat{\sigma}^-_{i+1}+\hat{\sigma}^-_i\hat{\sigma}^+_{i+1}\right).
\end{equation}
In the Hamiltonian $\hat{H}_S$ the constant $\epsilon$ denotes the
energy level of the spins and $\kappa$ the strength of the
spin-spin interaction.

The Hamiltonians $\hat{H}_{Bj}$ of the reservoirs coupled to the
first spin $(j=1)$ and the last spin $(j=3)$ are given by
\begin{equation}
\hat{H}_{Bj}=\sum_n \omega_{n,j}\hat{b}^\dag_{n,j}\hat{b}_{n,j}.
\end{equation}
The interaction between the spin subsystem and the bosonic baths
is described by
\begin{equation}
\hat{H}_{SBj}=\hat{\sigma}_{j}^{+}\sum_{n}g_{n}^{(j)}\hat{b}_{n,j}+\hat{\sigma}_{j}^{-}\sum_{n}g_{n}^{(j)*}\hat{b}_{n,j}^{\dag}.
\label{HSBj}
\end{equation}
Of course, $\hat{\sigma}_{j}^{\pm},\hat{\sigma}_{j}^{z}$ are the
well-known Pauli matrices and $\hat{b}^\dag_{n,j}$ and
$\hat{b}_{n,j}$ denote bosonic creation and annihilation
operators. The constants $\omega_{n,j}$ and $g_{n}^{(j)}$ denote
frequencies of the bosonic modes and amplitudes of the transitions
due to spin-boson interaction, respectively. In this paper units
are chosen such that $\hbar=k_B=1$.

In the Born-Markov approximation the dynamics of the reduced
density matrix $\hat{\rho}$ of the spin subsystem is described by
\cite{TOQS}:

\begin{equation}
\frac{d\hat{\rho}}{dt}=-i[\hat{H}_{S},\hat{\rho}]+{\cal
L}_{1}(\hat{\rho})+{\cal L}_{3}(\hat{\rho}),
\label{ME13}
\end{equation} with dissipators ($j=1$ and $j=3$)
\begin{eqnarray}
\mathcal{L}_{j}(\hat{\rho})&\equiv&\sum_{\mu,\nu}J_{\mu,\nu}^{(j)}(\omega_{j,\nu})\{[\hat{V}_{j,\mu},[\hat{V}_{j,\nu}^{\dag},\hat{\rho}]]
\\\nonumber
&
&-(1-e^{\beta_{j}\omega_{j,\nu}})[\hat{V}_{j,\mu},\hat{V}_{j,\nu}^{\dag}\hat{\rho}]\}.
\end{eqnarray}
The spectral density is given by
\begin{equation}
J_{\mu,\nu}^{(j)}(\omega_{j,\nu})=\int_{0}^{\infty}dse^{i\omega_{j,\nu}s}\langle
e^{-is\hat{H}_{B_j}}\hat{f}_{j,\nu}^{\dag}e^{is\hat{H}_{B_j}}\hat{f}_{j,\mu}\rangle_{j}.
\end{equation}
In the derivation of the master equation Eq. (\ref{ME13}) it is assumed that the interaction Hamiltonian Eq. (\ref{HSBj}) can be represented in the following way,
\begin{equation}
\hat{H}_{SBj}=\sum_{\mu}\hat{V}_{j,\mu}^\dag\hat{f}_{j,\mu}+\hat{V}_{j,\mu}\hat{f}_{j,\mu}^\dag,
\end{equation}
where the operators $\hat{V}_{j,\mu}$ and $\hat{f}_{j,\nu}$ describe
transitions in the spin and bath subsystems, respectively. Note,
that the index $j$ identifies the bath ($j=1$ and $j=3$) and the
index $\mu,\nu$ the number of the transition. As usual (cf. 
\cite{SinBurPet,Qui,Goldman,sergi}), the transition operators
$\hat{V}_{j,\mu}$ originate from the decomposition of the interaction Hamiltonian $\hat{H}_{SBj}$ into eigenoperators of the system Hamiltonian $\hat{H}_S$ and satisfy the following condition,
\begin{equation}
[\hat{H}_S,\hat{V}_{j,\mu}]=-\omega_{j,\mu}\hat{V}_{j,\mu}.
\label{commutator}
\end{equation}

To construct the explicit form of the transition operators we
need to find the eigenvectors and eigenvalues of the Hamiltonian
of the spin system $\hat{H}_S$. After straightforward
calculations we get a set of eigenvectors $|\lambda_i\rangle$,
where $(i=1,..,8)$:
\begin{equation}
|\lambda_1\rangle=|0,0,0\rangle,
\end{equation}
\begin{equation}
|\lambda_2\rangle=\frac{|0,0,1\rangle-|1,0,0\rangle}{\sqrt{2}},
\end{equation}
\begin{equation}
|\lambda_3\rangle=\frac{|0,1,1\rangle-|1,1,0\rangle}{\sqrt{2}},
\end{equation}
\begin{equation}
|\lambda_4\rangle=|1,1,1\rangle,
\end{equation}
\begin{equation}
|\lambda_5\rangle=\frac{|1,0,0\rangle-\sqrt{2}|0,1,0\rangle+|0,0,1\rangle}{2},
\end{equation}
\begin{equation}
|\lambda_6\rangle=\frac{|1,1,0\rangle-\sqrt{2}|1,0,1\rangle+|0,1,1\rangle}{2},
\end{equation}
\begin{equation}
|\lambda_7\rangle=\frac{|1,0,0\rangle+\sqrt{2}|0,1,0\rangle+|0,0,1\rangle}{2},
\end{equation}
\begin{equation}
|\lambda_8\rangle=\frac{|1,1,0\rangle+\sqrt{2}|1,0,1\rangle+|0,1,1\rangle}{2},
\end{equation}
with corresponding eigenvalues $\lambda_i$, where $(i=1,..,8)$:
\begin{equation}
\lambda_1=-\lambda_4=-\frac{3\epsilon}{2},
\end{equation}
\begin{equation}
\lambda_2=-\lambda_3=-\frac{\epsilon}{2},
\end{equation}
\begin{equation}
\lambda_5=-\lambda_8=-\frac{\epsilon}{2}-\sqrt{2}\kappa,
\end{equation}
\begin{equation}
\lambda_6=-\lambda_7=\frac{\epsilon}{2}-\sqrt{2}\kappa.
\end{equation}

Following \cite{Goldman} it is easy to see that in the case of
three spins the operators $\hat{V}_{j,\mu}$ have the form:
\begin{equation}
\hat{V}_{1,1}=\frac{1}{\sqrt{2}}\left(-|\lambda_1\rangle\langle\lambda_2|+|\lambda_3\rangle\langle\lambda_4|-|\lambda_5\rangle\langle\lambda_6|+|\lambda_7\rangle\langle\lambda_8|\right),
\end{equation}
\begin{equation}
\hat{V}_{1,2}=\frac{1}{2}\left(|\lambda_1\rangle\langle\lambda_5|-|\lambda_2\rangle\langle\lambda_6|-|\lambda_7\rangle\langle\lambda_3|+|\lambda_8\rangle\langle\lambda_4|\right),
\end{equation}
\begin{equation}
\hat{V}_{1,3}=\frac{1}{2}\left(|\lambda_1\rangle\langle\lambda_7|+|\lambda_2\rangle\langle\lambda_8|+|\lambda_5\rangle\langle\lambda_3|+|\lambda_6\rangle\langle\lambda_4|\right),
\end{equation}
\begin{equation}
\hat{V}_{3,1}=\frac{1}{\sqrt{2}}\left(|\lambda_1\rangle\langle\lambda_2|-|\lambda_3\rangle\langle\lambda_4|-|\lambda_5\rangle\langle\lambda_6|+|\lambda_7\rangle\langle\lambda_8|\right),
\end{equation}
\begin{equation}
\hat{V}_{3,2}=\frac{1}{2}\left(|\lambda_1\rangle\langle\lambda_5|+|\lambda_2\rangle\langle\lambda_6|+|\lambda_7\rangle\langle\lambda_3|+|\lambda_8\rangle\langle\lambda_4|\right),
\end{equation}
\begin{equation}
\hat{V}_{3,3}=\frac{1}{2}\left(|\lambda_1\rangle\langle\lambda_7|-|\lambda_2\rangle\langle\lambda_8|-|\lambda_5\rangle\langle\lambda_3|+|\lambda_6\rangle\langle\lambda_4|\right).
\end{equation}
The frequencies of the transitions are given by:
\begin{equation}
\omega_{1,1}\equiv\omega_{3,1}\equiv\omega_1=\epsilon,
\end{equation}
\begin{equation}
\omega_{1,2}\equiv\omega_{3,2}\equiv\omega_2=\epsilon-\sqrt{2}\kappa,
\end{equation}
\begin{equation}
\omega_{1,3}\equiv\omega_{3,3}\equiv\omega_3=\epsilon+\sqrt{2}\kappa.
\end{equation}
In following \cite{SinBurPet} we choose the bosonic bath as the
infinite set of harmonic oscillators and the coupling constants to
be frequency independent, so that
$J^{(j)}(\omega_{\nu})=\gamma_{j}n_{j}(\omega_{\nu})$, where
$n_{j}(\omega_{\nu})$ is the Bose distribution,
$n_{j}(\omega_{\nu})=(e^{\beta_{j}\omega_{\nu}}-1)^{-1}$. The dissipators $\mathcal{L}_j$ of the master equation Eq. (\ref{ME13}) take the following form,
\begin{eqnarray}
\label{diss}
\mathcal{L}_j(\hat{\rho})&=&\sum_{i=1}^3\gamma_j \left(n_j(\omega_i)+1\right)\left(\hat{V}_{j,i}\hat{\rho}\hat{V}_{j,i}^\dag-\frac{1}{2}[\hat{V}_{j,i}^\dag\hat{V}_{j,i},\hat{\rho}]_+\right)\\\nonumber& &+\gamma_j n_j(\omega_i)\left(\hat{V}_{j,i}^\dag\hat{\rho}\hat{V}_{j,i}-\frac{1}{2}[\hat{V}_{j,i}\hat{V}_{j,i}^\dag,\hat{\rho}]_+\right),
\end{eqnarray}
where $[A,B]_+\equiv AB+BA$ denotes the anticommutator.

In the basis of eigenvectors of the Hamiltonian $\hat{H}_S$ the
equation for the diagonal elements of the density matrix
$\hat{\rho}$ can be written as:
\begin{equation}
\frac{d}{dt}\left(\begin{array}{c}
\rho_{11}(t)\\
\rho_{22}(t)\\
\vdots\\
\rho_{88}(t)\end{array}\right)=B\left(\begin{array}{c}
\rho_{11}(t)\\
\rho_{22}(t)\\
\vdots\\
\rho_{88}(t)\end{array}\right),
\end{equation}
where $B$ is a $8\times8$ matrix of constant coefficients. The fact that the equation for the diagonal elements decouples from the non-diagonal ones is a consequence of the diagonal form of the semigroup generator, Eq. (\ref{diss}), and the diagonal form of the product of the transition operators $\hat{V}_{j,i}\hat{V}_{j,i}^\dag$. There
are 28 non-diagonal elements. The dynamical equations for them can
be divided in three groups. The first group consists of 4
non-diagonal elements the time dependence of which is trivial:
\begin{equation}
\rho_{i,j}(t)=\rho_{i,j}(0)e^{s_{i,j}t},
\end{equation}
where $s_{i,j}$ is a constant. The second group consists of 6
couples of non-diagonal elements which satisfy the following
system of equations:
\begin{equation}\label{M2}
\frac{d}{dt}\left(\begin{array}{c}
\rho_{i,j}(t)\\
\rho_{k,l}(t)\end{array}\right)=M^p_2\left(\begin{array}{c}
\rho_{i,j}(t)\\
\rho_{k,l}(t)\end{array}\right),
\end{equation}
where $M^p_2$ is a $2\times2$ matrix with constant coefficients.
The third group consists of 3 quadruples of non-diagonal elements,
each of them satisfies the following system of differential
equations:
\begin{equation}\label{M4}
\frac{d}{dt}\left(\begin{array}{c}
\rho_{i1,j1}(t)\\
\vdots\\
\rho_{i4,j4}(t)\end{array}\right)=M^p_4\left(\begin{array}{c}
\rho_{i1,j1}(t)\\
\vdots \\
\rho_{i4,j4}(t)\end{array}\right),
\end{equation}
where $M^p_4$ is a $4\times4$ matrix of constant coefficients.

\section{Analytical Solution}

The exact analytical solution for the diagonal elements in the
basis of eigenvectors $\{|\lambda_i\rangle\}$ of the Hamiltonian
$\hat{H}_S$ has the following form:
\begin{equation}
\left(\begin{array}{c}
\rho_{11}(t)\\
\rho_{22}(t)\\
\vdots\\
\rho_{88}(t)\end{array}\right)=RJ(t)R^{-1}\left(\begin{array}{c}
\rho_{11}(0)\\
\rho_{22}(0)\\
\vdots\\
\rho_{88}(0)\end{array}\right),
\end{equation}
where R is the nondegenerate matrix
\begin{equation}
R=\left(\begin{array}{cccccccc}
\frac{A^+C^+}{A^-C^-} & -\frac{C^+}{C^-} & \frac{A^+C^+}{A^-C^-} & -\frac{C^+}{C^-} & -\frac{A^+}{A^-} & 1 & -\frac{A^+}{A^-} & 1\\
\frac{C^+}{C^-} & \frac{C^+}{C^-} & \frac{C^+}{C^-} & \frac{C^+}{C^-} & -1 & -1 & -1 & -1\\
\frac{A^+B^-}{A^-B^+} & -\frac{B^-}{B^+} & -\frac{A^+}{A^-} & 1 & \frac{A^+B^-}{A^-B^+} & -\frac{B^-}{B^+} & -\frac{A^+}{A^-} & 1\\
\frac{B^-}{B^+} & \frac{B^-}{B^+} & -1 & -1 & \frac{B^-}{B^+} & \frac{B^-}{B^+} & -1 & -1\\
\frac{A^+B^-C^+}{A^-B^+C^-} & -\frac{B^-C^+}{B^+C^-} & -\frac{A^+C^+}{A^-C^-} & \frac{C^+}{C^-} & -\frac{A^+B^-}{A^-B^+} & \frac{B^-}{B^+} & \frac{A^+}{A^-} & -1\\
\frac{B^-C^+}{B^+C^-} & \frac{B^-C^+}{B^+C^-} & -\frac{C^+}{C^-} & -\frac{C^+}{C^-} & -\frac{B^-}{B^+} & -\frac{B^-}{B^+} & 1 & 1\\
\frac{A^+}{A^-} & -1 & \frac{A^+}{A^-} & -1 & \frac{A^+}{A^-} & -1 & \frac{A^+}{A^-} & -1\\
1 & 1 & 1 & 1 & 1 & 1 & 1 & 1\end{array}\right),
\end{equation}
and $J(t)$ is a diagonal matrix given by
\begin{equation}
{\small J(t)=\mathrm{diag}\left(1,e^{-At},e^{-Bt/2},e^{-\left(A+B/2\right)t},e^{-Ct/2},e^{-\left(A+C/2\right)t},e^{-\left(B+C\right)t/2},e^{-\left(A+B/2+C/2\right)t}\right),}
\end{equation}
where the coefficients $A^\pm,B^\pm,C^\pm$ are
\begin{equation}
A^{\pm}=J^{(1)}(\pm\omega_{1})+J^{(3)}(\pm\omega_{1}),\quad
A=A^++A^-,
\end{equation}
\begin{equation}
B^{\pm}=J^{(1)}(\pm\omega_{2})+J^{(3)}(\pm\omega_{2}),\quad
B=B^++B^-,
\end{equation}
\begin{equation}
C^{\pm}=J^{(1)}(\pm\omega_{3})+J^{(3)}(\pm\omega_{3}),\quad
C=C^++C^-.
\end{equation}
The time dependence of the non-diagonal elements of the first
group has the form:
\begin{equation}
\rho_{1,4}(t)=\rho_{1,4}(0)e^{-t\left(\frac{A}{2}+\frac{B+C}{4}-i3\epsilon\right)},
\end{equation}
\begin{equation}
\rho_{2,3}(t)=\rho_{2,3}(0)e^{-t\left(\frac{A}{2}+\frac{B+C}{4}-i\epsilon\right)},
\end{equation}
\begin{equation}
\rho_{5,8}(t)=\rho_{5,8}(0)e^{-t\left(\frac{A}{2}+\frac{B+C}{4}-i\epsilon-i2\sqrt{2}\kappa\right)},
\end{equation}
\begin{equation}
\rho_{6,7}(t)=\rho_{6,7}(0)e^{-t\left(\frac{A}{2}+\frac{B+C}{4}+i3\epsilon-i2\sqrt{2}\kappa\right)}.
\end{equation}
The first couple of the non-diagonal elements of the second group
(Eq. (\ref{M2}), $p=1$) satisfy the following equation:
\begin{equation}
\frac{d}{dt}\left(\begin{array}{c}
\rho_{2,5}(t)\\
\rho_{8,3}(t)\end{array}\right)=M^1_2\left(\begin{array}{c}
\rho_{2,5}(t)\\
\rho_{8,3}(t)\end{array}\right),
\end{equation}
where
\begin{equation}
M^1_2=\left(-\frac{A}{2}-\frac{B}{4}-i\sqrt{2}\kappa\right)+\frac{1}{2}\left(\begin{array}{cc}
-C^- & C^+\\
C^- & -C^+\end{array}\right).
\end{equation}
All the other couples of non-diagonal elements from the second
group satisfy equations of a similar kind. The explicit form of
the matrices $M^i_2$ ($i=2,..,6$) can be found in the Appendix.
The solution for the corresponding couple of non-diagonal elements
can be constructed with the help of the following formula:
\begin{eqnarray}
& &\exp{\left[t\Delta+t\left(\begin{array}{cc}
-\alpha & 0\\
0 & -\beta\end{array}\right)\pm t\left(\begin{array}{cc}
0 & \beta\\
\alpha & 0\end{array}\right)\right]}=\\\nonumber& &\frac{e^{\Delta t
}}{\alpha+\beta}\left(\begin{array}{cc}
\beta+\alpha e^{-(\alpha+\beta)t} & 0\\
0 & \alpha+\beta
e^{-(\alpha+\beta)t}\end{array}\right)\pm\frac{e^{\Delta t
}\left(1-e^{-(\alpha+\beta)t}\right)}{\alpha+\beta}\left(\begin{array}{cc}
0 & \beta \\
\alpha & 0\end{array}\right).
\label{exp2}
\end{eqnarray}
In the case $\alpha>0$, $\beta>0$ and $\mathrm{Re}[\Delta]<0$ it is obvious that
\begin{equation}
\lim_{t\rightarrow\infty}\exp{\left[t\Delta+t\left(\begin{array}{cc}
-\alpha & 0\\
0 & -\beta\end{array}\right)\pm t\left(\begin{array}{cc}
0 & \beta\\
\alpha & 0\end{array}\right)\right]}=0.
\end{equation}
This means that all non-diagonal elements of the second group will vanish at asymptotic times.

The first quadruple of the non-diagonal elements of the third
group (Eq. (\ref{M4}), $p=1$) satisfies the following system of
equations:
\begin{equation}
\frac{d}{dt}\left(\begin{array}{c}
\rho_{1,2}(t)\\
\rho_{3,4}(t)\\
\rho_{5,6}(t)\\
\rho_{7,8}(t)\end{array}\right)=M^1_4\left(\begin{array}{c}
\rho_{1,2}(t)\\
\rho_{3,4}(t)\\
\rho_{5,6}(t)\\
\rho_{7,8}(t)\end{array}\right),
\end{equation}
where $M^1_4$
\begin{equation}
M^1_4=-\frac{A}{2}+i\epsilon +T_4^1,
\end{equation}
where $T_4^1$ is the following matrix
\begin{equation}
T_4^1=\frac{1}{2}\left(\begin{array}{cccc}
-B^--C^- & 0 & -G^+ & H^+\\
0 & -B^+-C^+ & H^- & -G^-\\
-G^- & H^+ & -B^+-C^- & 0\\
H^- & -G^+ & 0 & -B^--C^+\end{array}\right)
\end{equation}
and the constants are
\begin{equation}
F^{\pm}=J^{(1)}(\pm\omega_{1})-J^{(3)}(\pm\omega_{1}),
\end{equation}
\begin{equation}
G^{\pm}=J^{(1)}(\pm\omega_{2})-J^{(3)}(\pm\omega_{2}),
\end{equation}
\begin{equation}
H^{\pm}=J^{(1)}(\pm\omega_{3})-J^{(3)}(\pm\omega_{3}).
\end{equation}
 The solution of the above system of equations has a complicated
form. Instead of presenting it we will show that the real part of
the maximum eigenvalues of the matrix $M^1_4$ is strictly
negative:
\begin{equation}
\mathrm{Re}\left[\lambda_{\mathrm{Max}}(M^1_4)\right]<0.
\end{equation}
This means that for long times the solution for this set of the
non-diagonal elements will converge to zero. In order to prove
that, we need first to analyze the eigenvalues of the matrix
$T_4^1$. After some straightforward calculation one can see that:
\begin{equation}
\lambda_{1,..,4}=\frac{1}{4}\left(-B-C\pm|B_G\pm C_H|\right),
\end{equation}
where the constant $B_G$ is
\begin{equation}
B_G=\sqrt{(B^+-B^-)^2+4G^+G^-}
\end{equation}
and the constant $C_H$ is
\begin{equation}
C_H=\sqrt{(C^+-C^-)^2+4H^+H^-}.
\end{equation}
Taking into account that $B$, $C$, $B_G$, $C_H>0$ the maximum
eigenvalue of the matrix $T_4^1$ is:
\begin{equation}
\lambda_{\mathrm{Max}}=\frac{1}{4}\left(-B-C+B_G+C_H\right).
\end{equation}
It is easy to see that the difference $B-B_G$ is non negative:
\begin{equation}
B\geq B_G\Leftrightarrow B^2\geq B^2_G
\end{equation}
\begin{equation}
\Leftrightarrow(B^++B^-)^2\geq (B^+-B^-)^2+4G^+G^-
\end{equation}
\begin{equation}
\Leftrightarrow B^+B^--G^+G^-\geq 0.
\end{equation}
Recalling the explicit expression for $B^\pm$ and $G^\pm$ it
follows:
\begin{eqnarray}
&B^+B^-&-G^+G^-=\\\nonumber & &
2J^{(1)}(\omega_{2})J^{(3)}(-\omega_{2})+2J^{(3)}(\omega_{2})J^{(1)}(-\omega_{2})\geq0.
\end{eqnarray}
In a similar way one can show that $C-C_H\geq0$. This means that
\begin{equation}
\mathrm{Re}\left[\lambda_{\mathrm{Max}}(M^1_2)\right]=-\frac{A}{2}-\frac{1}{4}\left(B+C-B_G-C_H\right)<0,
\end{equation}
i.e., the maximum value of the real part of the eigenvalues of the
matrix $M^1_4$ is negative and with time the corresponding
non-diagonal elements will converge to zero. A similar proof
holds for the other non-diagonal elements from the third group. An
explicit form of the corresponding matrices can be found in the
Appendix. More detailed calculations can be found in
\cite{Pumulo}.

From the above discussion it follows that all non-diagonal
elements converge to zero and a stationary state for the reduced
density matrix is given by the long time limit of the diagonal
elements, i.e.,
\begin{equation}
\lim_{t\rightarrow\infty}\rho_{ii}(t)=\frac{1}{ABC}\left(\begin{array}{c}
A^+B^+C^+\\
A^-B^+C^+\\
A^+B^-C^-\\
A^-B^-C^-\\
A^+B^-C^+\\
A^-B^-C^+\\
A^+B^+C^-\\
A^-B^+C^-\end{array}\right).
\end{equation}
As an example of the dynamics of the system, we consider an
initial $|W_3\rangle$ state for the three spin chain, i.e.,
\begin{equation}
|W_3\rangle=\frac{1}{\sqrt{3}}\left(|1,0,0\rangle+|0,1,0\rangle+|0,0,1\rangle\right).
\label{eq:w3}
\end{equation}
The only non-zero non-diagonal elements will be
$\langle\lambda_5|\hat{\rho}|\lambda_7\rangle$,
$\langle\lambda_6|\hat{\rho}|\lambda_8\rangle$ and their
transpositions,
\begin{eqnarray}
\langle\lambda_5|\hat{\rho}|\lambda_7\rangle&=&\langle\lambda_7|\hat{\rho}|\lambda_5\rangle^*\\\nonumber
&=&e^{-(B+C)t/4+i2\sqrt{2}\kappa t}\frac{A^++A^-e^{-At}}{6A},
\end{eqnarray}
and
\begin{eqnarray}
\langle\lambda_6|\hat{\rho}|\lambda_8\rangle&=&\langle\lambda_8|\hat{\rho}|\lambda_6\rangle^*\\\nonumber&=&-A^-e^{-(B+C)t/4+i2\sqrt{2}\kappa
t}\frac{1-e^{-At}}{6A}.
\end{eqnarray}
The diagonal elements have the following form
\begin{equation}
\langle\lambda_i|\hat{\rho}|\lambda_i\rangle=\frac{1}{ABC}\left(\frac{r_{i5}+r_{i7}}{2}+\frac{\sqrt{2}}{2}\left(r_{i7}-r_{i5}\right)\right), i=1,\dots, 8,
\end{equation}
where
\begin{equation}
r_{15}=g_A^+f_B^+g_C^+,\quad r_{17}=g_A^+g_B^+f_C^+,\quad r_{25}=f_A^-f_B^+g_C^+,\quad r_{27}=f_A^-g_B^+f_C^+,
\end{equation}
\begin{equation}
r_{35}=g_A^+g_B^-f_C^-,\quad r_{37}=g_A^+f_B^-g_C^-,\quad r_{45}=f_A^-g_B^-f_C^-,\quad r_{47}=f_A^-f_B^-g_C^-,
\end{equation}
\begin{equation}
r_{55}=g_A^+g_B^-g_C^+,\quad r_{57}=g_A^+f_B^-f_C^+,\quad r_{65}=f_A^-g_B^-g_C^+,\quad r_{67}=f_A^-f_B^-f_C^+,
\end{equation}
\begin{equation}
r_{75}=g_A^+f_B^+f_C^-,\quad r_{77}=g_A^+g_B^+g_C^-,\quad r_{85}=f_A^-f_B^+f_C^-,\quad r_{87}=f_A^-g_B^+g_C^-.
\end{equation}
The functions $f_{\{A,B,C\}}^\pm$ and $g_{\{A,B,C\}}^\pm$ are given by
\begin{equation}
f_{\{A,B,C\}}^\pm=\{A,B,C\}_\pm\left(1-e^{-\frac{t\{2A,B,C\}}{2}}\right)
\end{equation}
and
\begin{equation}
g_{\{A,B,C\}}^\pm=\left(\{A,B,C\}_\pm+\{A,B,C\}_\mp e^{-\frac{t\{2A,B,C\}}{2}}\right).
\end{equation}
For example, the expression for the $f_A^+$ or $g^-_B$ reads $f_A^+=A_+(1-e^{-At})$ and $g^-_B=B_-+B^+e^{-Bt/2}$, respectively.

\section{Results and Discussion}

In this section we analyze the entanglement between the first and
the third spin in the chain. The dynamics of the concurrence
\cite{woot} is presented in Fig. 1 and Fig. 2. Figure 1 shows
that for different temperatures of the baths the system behaves in
different ways. For $\beta_1=\beta_3=10$ one can see the well
known phenomena of sudden death and sudden birth of entanglement
\cite{eberly}. For other ranges of temperatures (curve (b)) one
can see that the system relatively quickly reaches the steady
state entanglement. For the lower temperatures of the baths (curve (a)) there is a competition between the unitary evolution and dissipation. During the short period of evolution, $\kappa t<10$, the oscillations of the concurrence due to the XX-interaction in the spin chain (time is measured in the dimensionless units $\kappa t$). However, due to the interaction with dissipative environments, this oscillation will disappear and after some time the thermal entanglement is created. In the case of higher temperatures of the baths (curve (b)) dissipative dynamics dominates over the unitary one.  Figure 2 illustrates the independence of the steady-state concurrence from the initial conditions; curve (a)
corresponds to a $|W_3\rangle$ initial state and one again can
see the sudden birth and death of entanglement, but after times
($\kappa t_{SS}$) of order $150$ the system converges to its steady
state and the concurrence remains constant; curve (b) depicts the
dynamics of the concurrence for the initial ``spin-up" state
$|1,1,1\rangle$ and one can see that the system converges to its
steady state relatively quickly ($\kappa t_{SS}\sim50$). In the case of the initial state $|W_3\rangle$ (curve (a)) one can see a behavior similar to the case of Figure 1, curve (a). However, the ``spin-up" initial state of the spin subsystem is not ``involved" in the XX-interaction (all spins are up, there is no exchange of excitations) and the system exponentially decays into a steady state, so that there is no oscillation of the concurrence.  

In the Figures 3 and 4 we compare the steady state concurrence for
a two qubit system (the corresponding expression is taken from
\cite{SinBurPet}) and the steady state concurrence between the
first and the third qubit for the three qubit system. In Fig. 3 we
analyze the difference between the steady-state concurrences in the equilibrium case
(both baths at the same temperature) as a function of the energy
of the spins and of the temperature of the baths. For temperatures
of the reservoirs $T<0.3$ we can see that the amount of steady-state concurrence is higher for a three-spin system (difference is positive)
than for a two-spin system for all energies of the spins $\epsilon$. Figure 4 addresses the
non-equilibrium case. One can see that in the range of
temperatures $T_1<0.5$ and $T_2<0.5$ the steady-state concurrence
for the three-spin system reaches higher values
than the two-spin one. But as in the symmetric two-spin case we observe that the steady-state entanglement reaches
a maximal value in the equilibrium case. This is the same behavior
that Huang \cite{Huang} had found in the numerical study of a
slightly more complicated model. But the beauty of our approach
is that we have found an exact analytical expression for the reduced density
matrix showing the same behavior and prove the existence of the
non-equilibrium steady state.

In conclusion, we found an analytical expression for a three qubit
spin chain coupled to bosonic baths at different temperatures. We
studied the dynamics of the system and showed its convergence to a
steady state. In the range of parameters $\epsilon\sim\kappa>>\gamma$ that we investigated the steady state of the three-spin chain has a diagonal form in the basis of the eigenstates of the Hamiltonian of the system $H_S$. This diagonal form implies the absence of a heat current between the spins in the steady state. In the context of non-equilibrium quantum transport a similar behavior was discussed by T. Prosen and B. \ifmmode \check{Z}\else \v{Z}\fi{}unkovi\ifmmode \check{c}\else \v{c}\fi{} in \cite{Prosen}. We analyzed the dynamics of entanglement and
performed a comparison of the steady state concurrence of two and
three spin chains. We found a range of parameters in which the
three spin chain contains more quantum correlations in the steady
state than two spin one. Similarity in the behavior of the steady
state concurrence for two and three spin chain motivate us to
extend our research for longer chains and more complicated
configurations  of the spins.

\begin{figure}
\begin{center}
\includegraphics[width=0.9\linewidth]{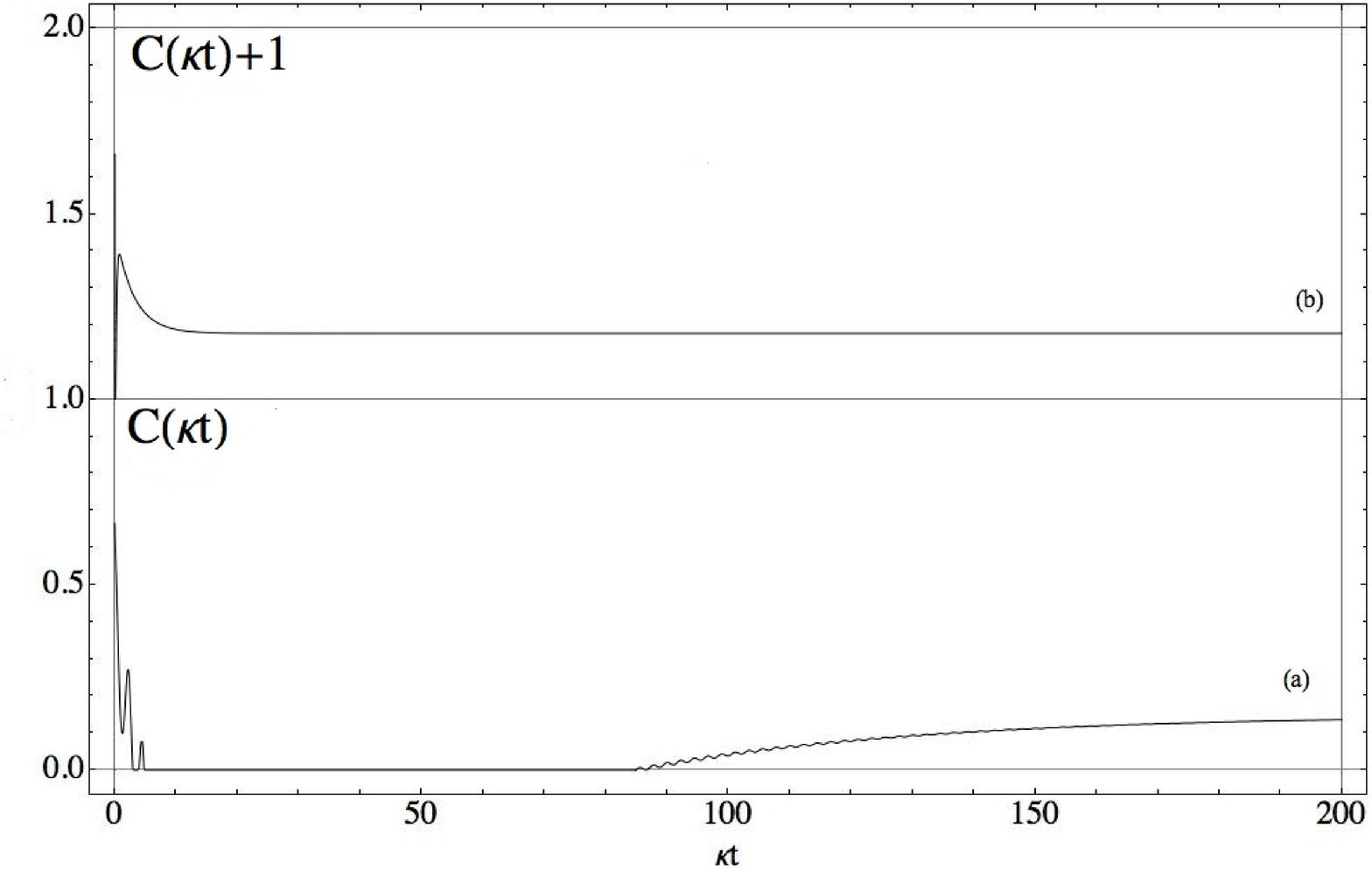}
\caption{Concurrence as a function of the dimensionless parameter $\kappa t$ for different temperatures
of the heat baths. The initial state of the three qubit system is the
$|W_3\rangle$ state defined in Eq.(\ref{eq:w3}). Curve $(a)$ corresponds to
$\beta_1=\beta_3=10$, curve $(b)$ to $\beta_1=5$ and $\beta_3=1$; all
the other parameters are the same:
$\epsilon=3/2$,$\kappa=1$,$\gamma_1=\gamma_3=1/50$.}\label{fig1}
\end{center}
\end{figure}

\begin{figure}
\begin{center}
\includegraphics[width=0.9\linewidth]{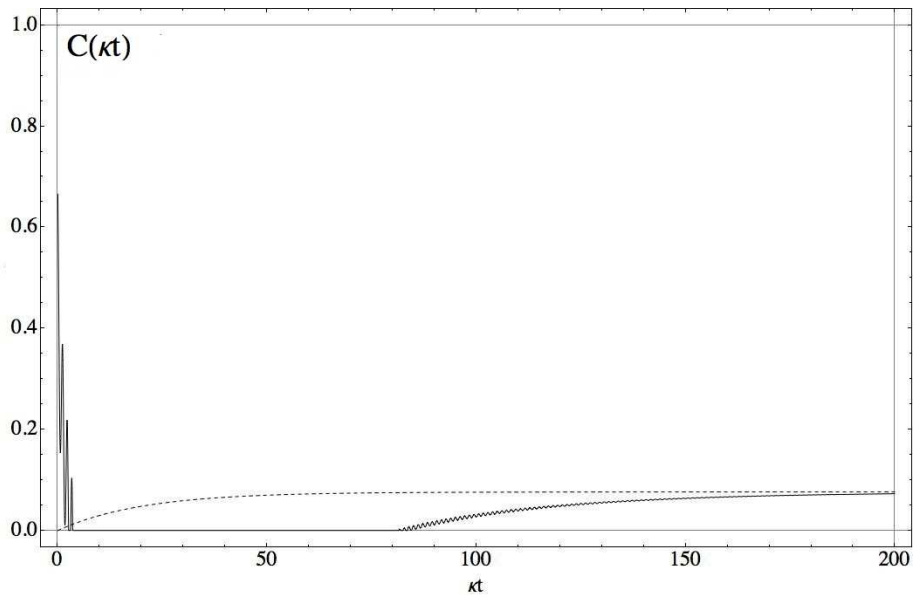}
\caption{Concurrence as a function of the dimensionless parameter $\kappa t$ for different initial states of the
three qubit system. The solid curve corresponds to the $|W_3\rangle$
state and the dashed curve corresponds to factorized spin-up state
$|\psi_0\rangle=|1,1,1\rangle$; all the other parameters are the
same:
$\epsilon=3$, $\kappa=2$, $\gamma_1=\gamma_3=1/20$, $\beta_1=10$ and $\beta_3=15$.}\label{fig2}
\end{center}
\end{figure}

\begin{figure}
\begin{center}
\includegraphics[width=0.9\linewidth]{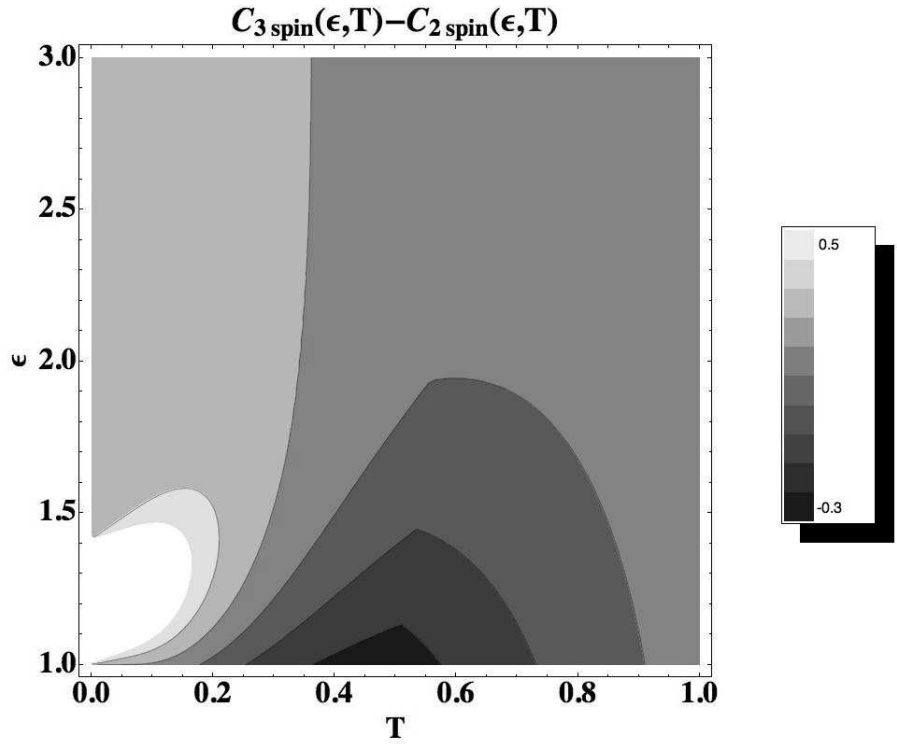}
\caption{Difference of the steady-state concurrence between the first and last spin of a three-spin chain ($C_{3\;\mathrm{ spin}}(\epsilon,T)$)  and the spins of a two-spin chain ($C_{2\;\mathrm{ spin}}(\epsilon,T)$) in thermal equilibrium as a function of the energy of the spins $\epsilon$ and temperature of the baths $T$. For both chains the parameters are $\gamma_1=\gamma_3=1/20$ for the first and the last spin, respectively, and $\kappa=1$.}\label{fig3}
\end{center}
\end{figure}

\begin{figure}
\begin{center}
\includegraphics[width=0.9\linewidth]{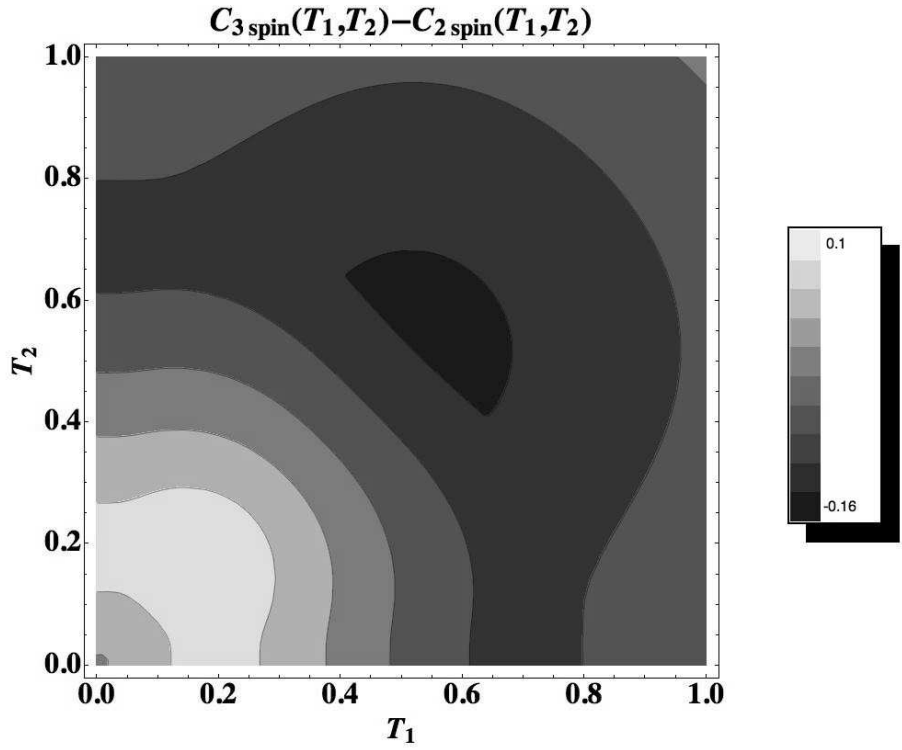}
\caption{Difference of the steady-state concurrence between the first and last spin of a three-spin chain ($C_{3\;\mathrm{ spin}}(T_1,T_2)$)
and the spins of a two-spin chain ($C_{2\;\mathrm{ spin}}(T_1,T_2)$) as a function of the temperatures of the
baths $T_1$ and $T_2$. For both chains the parameters are $\gamma_1=\gamma_3=1/20$ for the first and the last spin, respectively, $\kappa=1$ and $\epsilon=1.6$.}\label{fig4}
\end{center}
\end{figure}
\section*{Acknowledgements}
This work is based upon research supported by the South African
Research Chair Initiative of the Department of Science and
Technology and National Research Foundation.
\appendix
\section{Non-diagonal elements of the second and third groups}

The non-diagonal elements of the second group satisfy the
following equations:
\begin{equation}
\frac{d}{dt}\left(\begin{array}{c}
\rho_{1,3}(t)\\
\rho_{2,4}(t)\end{array}\right)=M^2_2\left(\begin{array}{c}
\rho_{1,3}(t)\\
\rho_{2,4}(t)\end{array}\right),
\end{equation}
where
\begin{equation}
M^2_2=\left(-\frac{B+C}{4}+i2\epsilon\right)+\left(\begin{array}{cc}
-A^- & -A^+\\
-A^- & -A^+\end{array}\right);
\end{equation}
and
\begin{equation}
\frac{d}{dt}\left(\begin{array}{c}
\rho_{1,6}(t)\\
\rho_{7,4}(t)\end{array}\right)=M^3_2\left(\begin{array}{c}
\rho_{1,6}(t)\\
\rho_{7,4}(t)\end{array}\right),
\end{equation}
where
\begin{equation}
M^3_2=\left(-\frac{A}{2}-\frac{B}{4}+i2\epsilon-i2\sqrt{2}\kappa\right)+\frac{1}{2}\left(\begin{array}{cc}
-C^- & C^+\\
C^- & -C^+\end{array}\right);
\end{equation}
and
\begin{equation}
\frac{d}{dt}\left(\begin{array}{c}
\rho_{1,8}(t)\\
\rho_{5,4}(t)\end{array}\right)=M^4_2\left(\begin{array}{c}
\rho_{1,8}(t)\\
\rho_{5,4}(t)\end{array}\right),
\end{equation}
where
\begin{equation}
M^4_2=\left(-\frac{A}{2}-\frac{C}{4}+i2\epsilon+i\sqrt{2}\kappa\right)+\frac{1}{2}\left(\begin{array}{cc}
-B^- & B^+\\
B^- & -B^+\end{array}\right);
\end{equation}
and
\begin{equation}
\frac{d}{dt}\left(\begin{array}{c}
\rho_{2,7}(t)\\
\rho_{6,3}(t)\end{array}\right)=M^5_2\left(\begin{array}{c}
\rho_{2,7}(t)\\
\rho_{6,3}(t)\end{array}\right),
\end{equation}
where
\begin{equation}
M^5_2=\left(-\frac{A}{2}-\frac{C}{4}+i\sqrt{2}\kappa\right)+\frac{1}{2}\left(\begin{array}{cc}
-B^- & B^+\\
B^- & -B^+\end{array}\right);
\end{equation}
and
\begin{equation}
\frac{d}{dt}\left(\begin{array}{c}
\rho_{5,7}(t)\\
\rho_{6,8}(t)\end{array}\right)=M^6_2\left(\begin{array}{c}
\rho_{5,7}(t)\\
\rho_{6,8}(t)\end{array}\right),
\end{equation}
where
\begin{equation}
M^6_2=\left(-\frac{B+C}{4}+i2\sqrt{2}\kappa\right)+\left(\begin{array}{cc}
-A^- & -A^+\\
-A^- & -A^+\end{array}\right).
\end{equation}

The non-diagonal elements of the third group satisfies the
following equations:

\begin{equation}
\frac{d}{dt}\left(\begin{array}{c}
\rho_{1,5}(t)\\
\rho_{2,6}(t)\\
\rho_{7,3}(t)\\
\rho_{8,4}(t)\end{array}\right)=M^2_4\left(\begin{array}{c}
\rho_{1,5}(t)\\
\rho_{2,6}(t)\\
\rho_{7,3}(t)\\
\rho_{8,4}(t)\end{array}\right),
\end{equation}
where $M^2_4$
\begin{eqnarray}
M^2_4&=&-\frac{B}{4}+i\epsilon-i\sqrt{2}\kappa+\\& &
\frac{1}{2}\left(\begin{array}{cccc}
-2A^--C^- & 2F^+ & H^+ & 0\\
2F^- & -2A^+-C^- & 0 & H^+\\
H^- & 0 & -2A^--C^+ & F^+\\
0 & H^- & 2F^- & -2A^+-C^+\end{array}\right);\nonumber
\end{eqnarray}
and
\begin{equation}
\frac{d}{dt}\left(\begin{array}{c}
\rho_{1,7}(t)\\
\rho_{2,8}(t)\\
\rho_{5,3}(t)\\
\rho_{6,4}(t)\end{array}\right)=M^3_4\left(\begin{array}{c}
\rho_{1,7}(t)\\
\rho_{2,8}(t)\\
\rho_{5,3}(t)\\
\rho_{6,4}(t)\end{array}\right),
\end{equation}
where $M^3_4$
\begin{eqnarray}
M^3_4&=&-\frac{C}{4}+i\epsilon+i\sqrt{2}\kappa+\\& &
\frac{1}{2}\left(\begin{array}{cccc}
-2A^--B^- & -2F^+ & -G^+ & 0\\
-2F^- & -2A^+-B^- & 0 & -G^+\\
-G^- & 0 & -2A^--B^+ & -F^+\\
0 & -G^- & -2F^- & -2A^+-B^+\end{array}\right).\nonumber
\end{eqnarray}

\bibliographystyle{elsarticle-num}
\bibliography{<your-bib-database>}

\begin{thebibliography}{00}

\bibitem{TOQS} H.-B. Breuer, F. Petruccione, {\it The Theory of Open
Quantum Systems}  (Oxford University Press, 2002)

\bibitem{braun} M.B. Plenio and S.F. Huelga, Phys. Rev. Lett., 88, 197901 (2002); D. Braun, {\it ibid}. 89, 277901
(2002); S. Diehl et al., Nature Physics, 4, 878 (2008).

\bibitem{TE} M.C. Arnesen et al., Phys. Rev. Lett. 87, 017901 (2001); X.G. Wang, Phys. Rev. A 64, 012313 (2001);
X.G. Wang et al., J. Phys. A 34, 11307 (2001); X.G. Wang,
 Phys. Rev. A 66, 034302 (2002); X.G. Wang, {\it ibid.} 66, 044305 (2002); G.L. Kamta and A.F. Starace,
Phys. Rev. Lett. 88, 107901 (2002); L. Zhou et al., Phys.
Rev. A 68, 024301 (2003); Y. Sun et al., ibid.
68, 044301 (2003); M. Cao and S. Zhu, {\it ibid.} 71,
034311 (2005).

\bibitem{SinBurPet} I. Sinaysky, F. Petruccione, D. Burgarth, Phys. Rev. A 78, 062301
(2008).

\bibitem{Huang} X.L. Huang, J.L. Guo, X.X. Yi, Phys. Rev. A 80, 054301,
(2009).

\bibitem{Qui} L. Quiroga, F.J. Rodriguez, M.E. Ramirez, R. Paris, Phys. Rev. A 75, 032308
(2007).

\bibitem{FL} C. Mej\'ia-Monasterio, T. Prosen and G. Casati, Europhys. Lett. 72, 520 (2005); T. Prosen and I. Pi\ifmmode \check{z}\else \v{z}\fi{}orn Phys. Rev. Lett. 101, 105701 (2008); T. Prosen and M. \ifmmode \check{Z}\else \v{Z}\fi{}nidari\ifmmode \check{c}\else \v{c}\fi{}, Phys. Rev. Lett. 105, 060603 (2010).

\bibitem{Goldman} M. Goldman, J. Magn. Reson. 149, 160 (2001).

\bibitem{sergi} A. Sergi, I. Sinayskiy, F. Petruccione, Phys. Rev. A 80, 012108,
(2009).

\bibitem{Pumulo} N. Pumulo, "Simple qubit systems in bosonic baths``, MSc Thesis, University of KwaZulu-Natal, (2011)

\bibitem{woot} W.K. Wootters, Phys. Rev. Lett. 80, 2245 (1998)

\bibitem{eberly} T. Yu and J.H. Eberly, Phys. Rev. Lett. 93, 140404
(2004)
\bibitem{Prosen} T. Prosen and B. \ifmmode \check{Z}\else \v{Z}\fi{}unkovi\ifmmode \check{c}\else \v{c}\fi{}, New J. Phys. 12, 025016 (2010)
\end{thebibliography}

\end{document}